\documentclass[graybox]{svmult}
\usepackage{bookmark}
    \makeatletter
    \def\toclevel@title{-1}
    \def\toclevel@author{0}
    \makeatother
\usepackage{newtxtext}
\usepackage[varvw]{newtxmath}
\usepackage{hyperref}       
\usepackage{makeidx}
\usepackage{multicol}
\usepackage[bottom]{footmisc}
\usepackage{graphicx}
    \graphicspath{{images/}}
\usepackage{tikz}
    \usetikzlibrary{calc, fadings}
    \tikzstyle{labl} = [text=white, inner sep=1mm, font={\Large}]
\usepackage{xcolor}
    \definecolor{commentgreen}{RGB}{34,139,34}
    \definecolor{stringpurple}{RGB}{160,32,240}
    \definecolor{keywordblue}{RGB}{0,0,255}

\usepackage[outline]{contour} 
    \newcommand{\bkc}[1]{\contour{black}{#1}} 

\usepackage{biblatex}
\usepackage{wasysym}

\newcommand{\uns}{\textunderscore}
\renewcommand{\vec}[1]{\boldsymbol{#1}} 

\newcommand{\rev}[1]{{#1}}

\begin{document}
\title*{STEMPO -- dynamic X-ray tomography phantom}

\author{Tommi Heikkil\"a} 
\institute{Tommi Heikkil\"a \at Unversity of Helsinki, Department of Mathematics and Statistics. \\
Pietari Kalmin Katu 5, 00560 Helsinki, Finland, \\
\email{tommi.heikkila@helsinki.fi}}

\maketitle

\abstract{This is the documentation for the Spatio-TEmporal Motor-POwered (STEMPO) phantom for dynamic X-ray tomography. Different measurements designed for testing dynamic tomography reconstruction algorithms are provided. The measured data and additional materials are available at \url{http://www.fips.fi/dataset.php} and \href{https://doi.org/10.5281/zenodo.7081688}{Zenodo} for open use by the scientific community, as long as the data and this documentation are appropriately referenced. \\[1em]
The files contain: (1) measurement data (\emph{sinograms}) from one static scan and two dynamic scans (same movement but different choice of projection directions). All data is provided both for 2D and 3D geometry in varying resolution levels; (2) approximation of the ground truth obtained using the known motion model; and (3) short example codes to showcase how the data could be used to test and validate dynamic tomography algorithms.}


\section{Introduction}
\begin{figure}[t!]
    \centering
    \includegraphics[width=0.9\columnwidth]{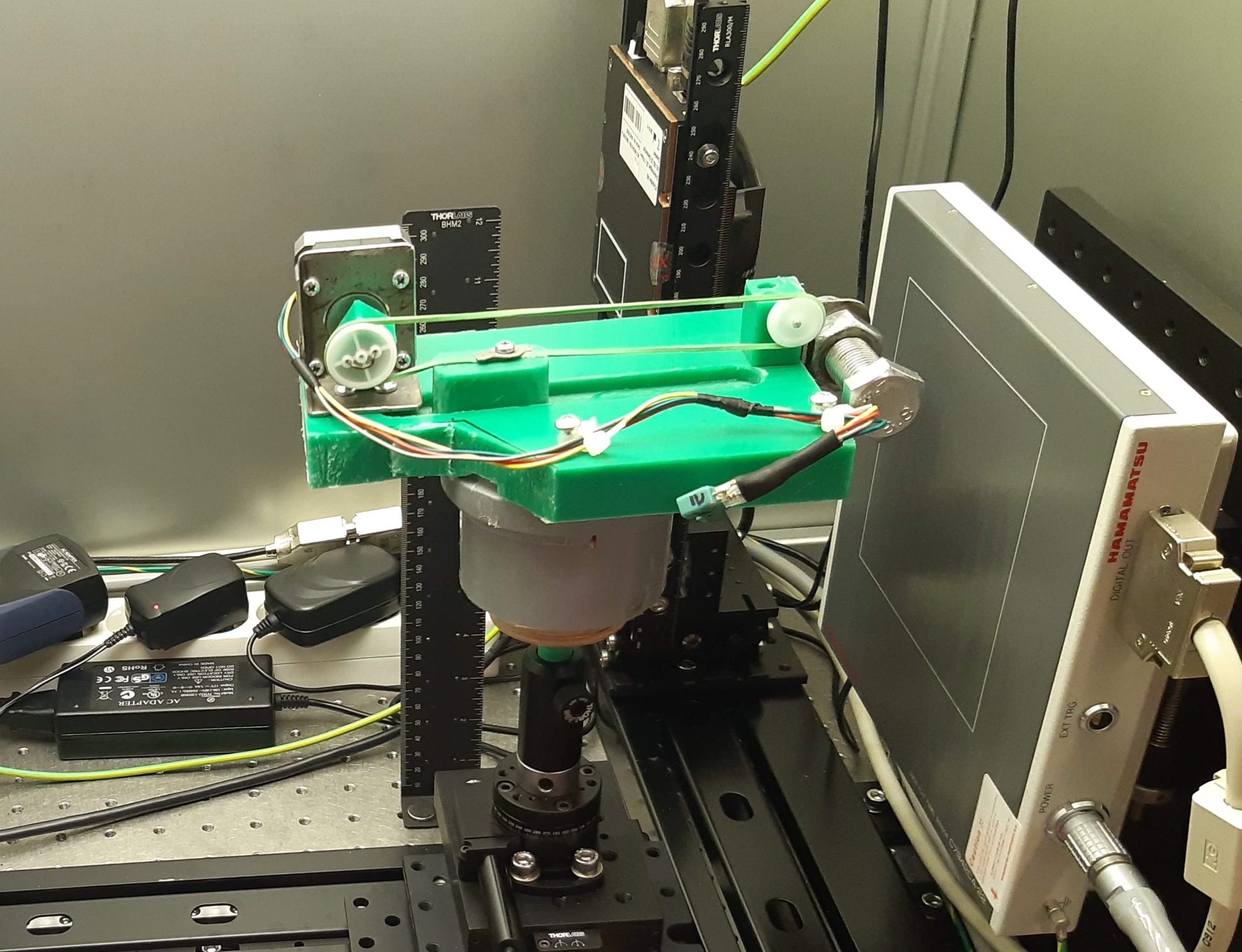}
    \caption{STEMPO phantom inside the micro-CT device. The movement mechanism including a stepper motor, a pulley system and a sled travelling in a straight cut channel are housed on top of the phantom. The moving block and the static object inside the body of the phantom (grey pipe) are below the lid and not visible from the outside, but they can be seen in Fig.~\ref{fig:phantomDetails}.}
    \label{fig:phantom}
\end{figure}

\begin{figure}[t!]
    \centering
    \begin{tikzpicture}
        \node (side) at (0,3.43) {\includegraphics[width=111mm]{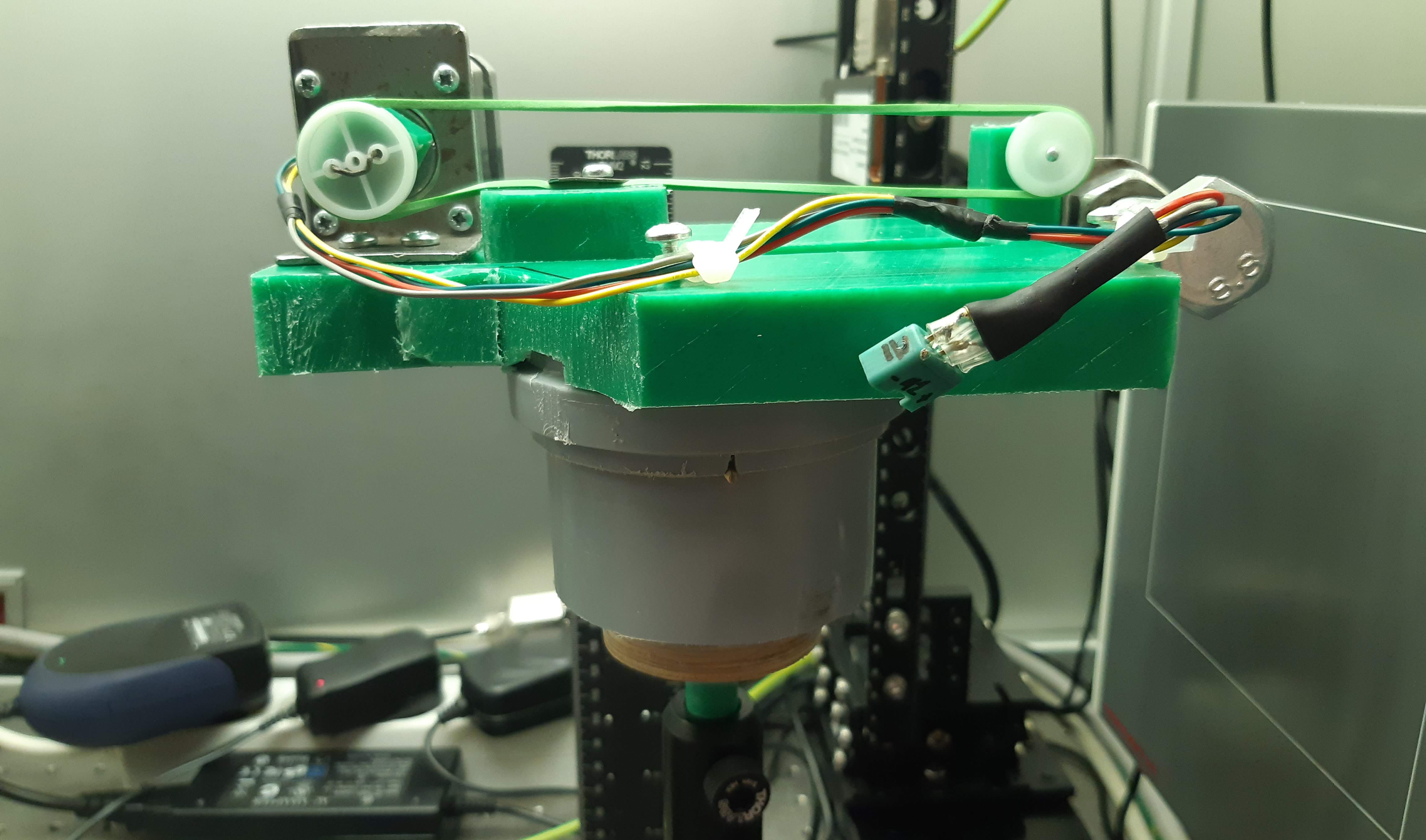}};
        \node (bottom) at (-2.8cm,-2) {\includegraphics[width=55mm]{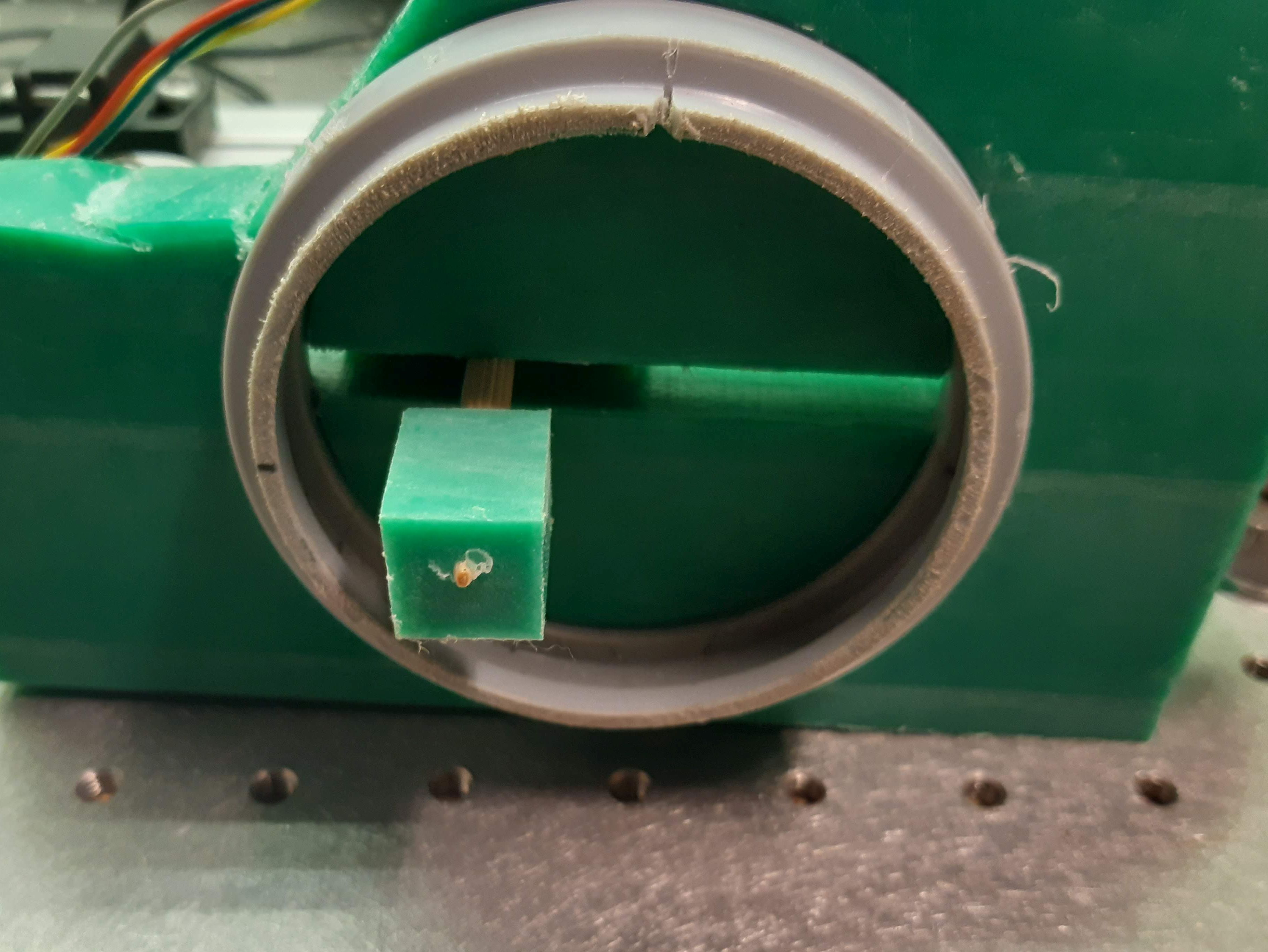}};
        \node (static) at (2.8cm,-2) {\includegraphics[width=55mm]{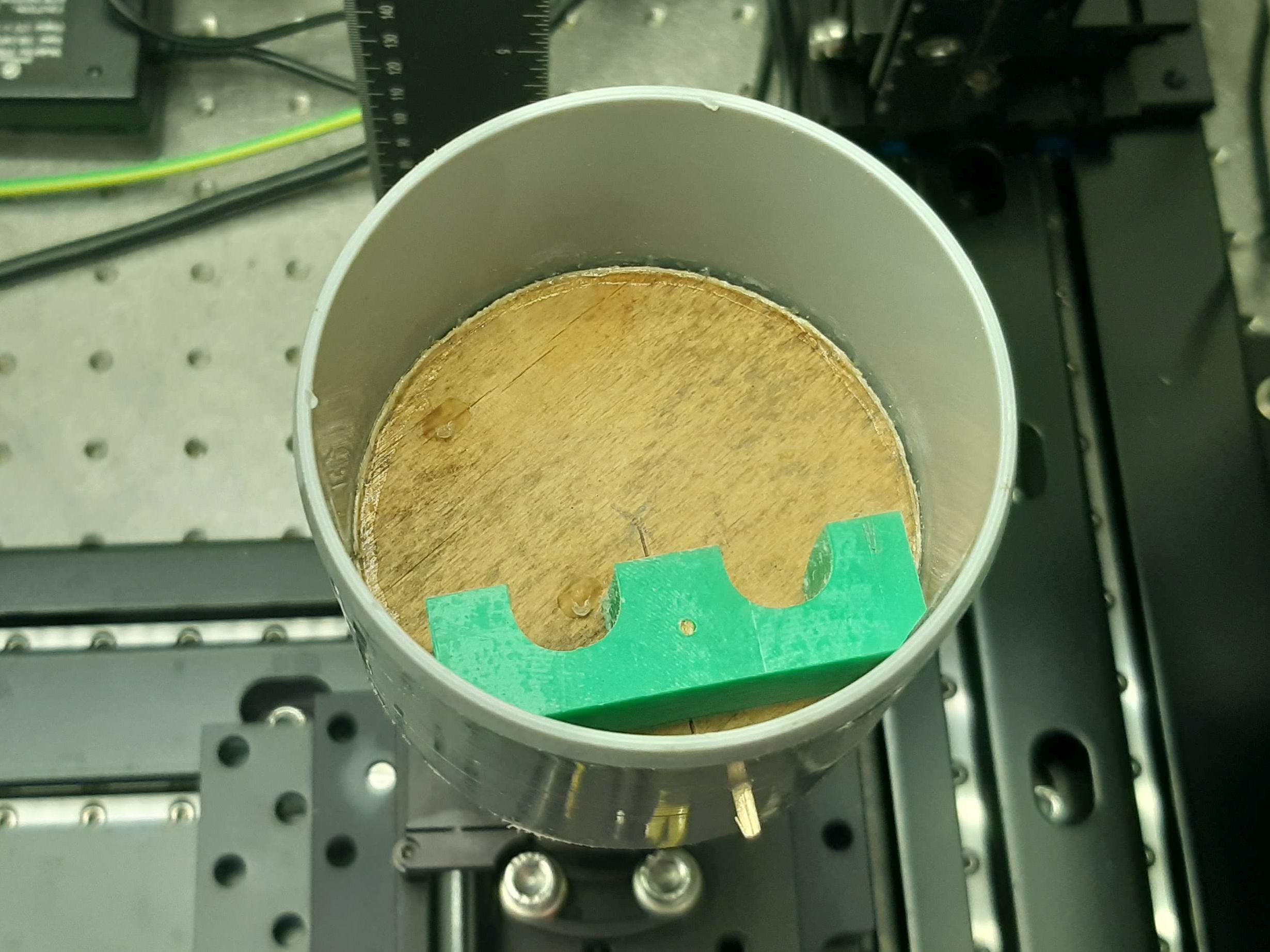}};

        \node[labl] (lid1) at (0,4) {\bkc{a}}; 
        \node[labl] (lid2) at (-5,-2) {\bkc{a}}; 
        \node[labl] (motor) at (-0.5,6.3) {\bkc{e}}; 
        \node[labl] (pipe1) at (0,2.5) {\bkc{b}}; 
        \node[labl] (pipe2) at (2.8,-0.7) {\bkc{b}}; 
        \node[labl] (block) at (-4.8,-3.5) {\bkc{c}}; 
        \node[labl] (static) at (4.8,-3.5) {\bkc{d}}; 
        \node[labl] (pulleys) at (0,5.6) {\bkc{f}}; 
        \node[labl] (sled) at (-1.05,4.95) {\bkc{g}}; 
        \node[labl] (weight) at (3.9,4.6) {\bkc{h}}; 
        \node[labl] (power) at (2.25,4.15) {\bkc{i}}; 
        \node[labl] (base) at (0,1.4) {\bkc{j}}; 

        \draw[white,thick] (block) -- (-3.7,-2.5);
        \draw[white,thick] (static) -- (3.3,-2.7);
        \draw[white,thick] (motor) -- (-1.6,6.2);
        \draw[white,thick] (pulleys) -- (-2.2,5.5);
        \draw[white,thick] (pulleys) -- (2.2,5.5);
    \end{tikzpicture}
    \caption{Key details of the STEMPO phantom with labels for a: \emph{lid}, b: \emph{bounding pipe}, c: \emph{moving block}, d: \emph{static object}, e: \emph{stepper motor}, f: \emph{pulleys}, g: \emph{sled}, h: \emph{counter weight}, i: \emph{power cable} and \emph{connector} and j: \emph{base} and \emph{attachment rod}. Top: side view of the whole phantom inside the CT device. In the bottom images the lid has been detached from the pipe for better visuals. On left: the lid and moving block viewed from below. On right: the pipe and the static object viewed from above. Only the components inside the pipe (i.e. parts b,c and d) are visible in the X-ray projections.}
    \label{fig:phantomDetails}
\end{figure}

The Spatio-TEmporal Motor-POwered (STEMPO) phantom is designed for collecting high resolution X-ray tomography images from a simple known target undergoing simple translation motion. It is a self-built phantom made of mostly high-density polyethylene (HDPE) plastic with a computer controlled stepper motor, which can move part of the phantom while the projections are being measured. The phantom is shown in Fig.~\ref{fig:phantom} and the key features are labelled and shown in Fig.~\ref{fig:phantomDetails}.

The goal of this contribution is to help (the mathematical community in particular) validate and compare novel and experimental reconstruction methods for dynamic tomography. Currently, openly available real data often comes from medical sources (cf. \cite{vandemeulebroucke2007} and \href{http://www.vmip.org/7.html}{VMIP.org}), where the temporal resolution is mostly coarse \cite{purisha2018} or the measurements come from specialized setups \cite{offenwert2020, heikkila2020} whereas some interesting reconstruction methods place limitations on the underlying movement or sampling \cite{burger2017, blanke2020} which can make the qualitative comparison \cite{hauptmann2021} of different methods difficult. 

STEMPO is designed to be a structurally simple and robust object with well controlled dynamic behaviour which can be repeated or tailored to different measurement setups and reconstruction settings. This ensures that collecting the data is relatively easy for the operator, a wide range of users and applications can be covered -- either by the current data or data collected in the future -- and all of the different data sets remain similar enough to allow reasonable comparison of different methods.

The paper is organized as follows. Section~\ref{sec:phantom} explains the structure and physical properties of the STEMPO phantom in detail. This motivates the different measurements and test algorithms. The measurement data and additional files are listed in section \ref{sec:contents} including a brief explanation of the typical mathematical model and some instructions for the users. 
The measurement setup and the projection geometry are presented in section~\ref{sec:measurements}. A couple of well-known example algorithms are briefly demonstrated in section~\ref{sec:examples} to illustrate how the data could be used. Finally since the data presented here is just a limited sample of the possible (and interesting) measurement setups, in section~\ref{sec:development} some future developments are discussed. However it is expected that this documentation remain valid even when new data is collected and included in Zenodo \cite{stempo}.

\section{Phantom structure} \label{sec:phantom}
The main body of the phantom is a short piece of grey polypropene (PP) pipe (height: 56 mm, outer diameter: 80 mm) which bounds the possible movement and contains a removable static object (roughly $54 \times 15 \times 20$ mm with some cutouts) made of green HDPE. The exact shape of the interior parts are shown in the two bottom images of Fig.~\ref{fig:phantomDetails} and also in the 3D reconstruction in Fig.~\ref{fig:fdkRecn}. \rev{Held in place by the pipe} is a flat HDPE lid which contains the stepper motor and the channel for the motor actuated sled to travel along. The \rev{moving HDPE  block ($15 \times 15 \times 20$ mm)} is fixed to the underside of the sled such that the block is visible on the imaging plane at similar height as the static object. The base of the phantom is rigidly attached to a round 12 mm HDPE rod which can be attached to the rotating platform.

The X-ray detector and sample rotator are controlled by a main computer. During the measurement process this computer sends additional messages to a \href{https://www.raspberrypi.org/}{Raspberry Pi} 3 model B which uses an Adafruit Stepper Motor HAT to translate these messages into commands for a Nema-17 stepper motor (model XY42STH34-0354A), with 200 full steps per revolution. Finer step length is possible using interleaved or microstepping (for 2 or 16 times the resolution respectively). The Raspberry Pi and the phantom are connected via a power cable (partially visible in Figs.~\ref{fig:phantom} and \ref{fig:phantomDetails}. Finally the stepper motor moves the object of interest via an elastic pulley. The diameter of the drive pulley is 23 mm which means that the minimum translation of the block is roughly $\pi \tfrac{23 \text{mm}}{200} \approx 0.36$ mm when using full steps. The whole process is automated and the user needs just to define the step length taken between each measured projection. Therefore the block can be moved to any position along the straight path for each projection.

\section{Data set contents} \label{sec:contents}

The data set contains the following files for MATLAB (here \texttt{b*} denotes binning or downsampling factor of 4, 8, 16 or 32):

\begin{itemize}
    \item[]\textbf{Data -- vol. 1}
    \item \texttt{stempo\uns static\uns 2d\uns b*.mat}
    \item \texttt{stempo\uns static\uns 3d\uns b*.mat}
    \item \texttt{stempo\uns cont360\uns 2d\uns b*.mat}
    \item \texttt{stempo\uns cont360\uns 3d\uns b*.mat}
    \item \texttt{stempo\uns seq8x45\uns 2d\uns b*.mat}
    \item \texttt{stempo\uns seq8x45\uns 3d\uns b*.mat}\\[.1em]
    \item[]\textbf{Additional files}
    \item \texttt{stempo\uns ground\uns truth\uns 2d\uns b4.mat}\\[.1em]
    \item[]\textbf{Example algorithms}
    \item \texttt{stempo\uns fbp\uns example.m}
    \item \texttt{stempo\uns fdk\uns example.m} (CUDA only)
    \item \texttt{stempo\uns pdfp\uns wavelet\uns 2d\uns example.m}
    \item \texttt{stempo\uns LplusS\uns 2d\uns example.m}
\end{itemize}

The intended (and most straightforward) use of these files requires \href{https://www.astra-toolbox.com/}{the ASTRA Toolbox} \cite{van2015astra, van2016fast} for implementing the forward operator, \href{https://www.cs.ubc.ca/labs/scl/spot/}{the Spot Linear-Operator Toolbox} \cite{spot} for more efficient computations and \href{https://github.com/Diagonalizable/HelTomo}{the HelTomo Toolbox} \cite{heltomo} for easy-to-use tools and functions for handling the data and initializing the operators. However other options (such as different programming languages) are possible as long as the measurement geometry is respected.

\subsection{Sinograms}
All of the data files contain a MATLAB structure array which includes the necessary metadata and information about the measurement geometry, and the \texttt{sinogram}. For the 2D geometry it is a numerical $P \times D$ array containing all of the measurements from $P = 360$ projections and $D$ detector elements, depending on the binning factor: $D = \tfrac{2240}{\text{binning}}$. For the 3D geometry the sinogram is a numerical $D_c \times P \times D_r$ array containing $P = 360$ projections with $D_r$ rows and $D_c$ columns, depending again on the binning factor: $D_r = \tfrac{2368}{\text{binning}}$ and $D_c = \tfrac{2240}{\text{binning}}$.

The sinograms are obtained by considering the logarithm of the ratio between the initial intensity and the measured intensity. The initial intensity for each individual projection is calculated by taking the mean of a $384 \times224$ pixel area known to be outside the object's X-ray projection. For computational efficiency the measurements have been binned by a factor of 4, 8, 16 or 32 before being organized into sinograms.

Binning is a standard procedure to lower the data resolution and improving the signal-to-noise ratio by summing the measured intensity values from multiple neighbouring detector elements, such as non-overlapping $2\times 2$ neighbourhoods (binning factor of $2$). This is also taken into account with the parameters such as relative pixel size. In preparing this data set, binning was applied to the raw projection images both horizontally and vertically to preserve the square shape of the detector elements even for the 2D sinograms.

If a smaller number of projections is needed, for example for sparse angle tomography, the sinograms can be downsampled with the function \texttt{subsample\uns sinogram.m}. This will produce a new tomographic measurement data structure which matches the projection directions given by the user.

For \texttt{stempo\uns static\uns *} and \texttt{stempo\uns cont360\uns *} (continuous) data sets the angular sampling is $360$ projections with $1^\circ$ angle interval between consecutive projections for just one full rotation around the phantom during the full movement cycle. This is the usual sampling strategy used in computed tomography.

For \texttt{stempo\uns seq8x45\uns *} (sequence) data sets the angular sampling is 360 projections with $8^\circ$ angle interval for a total of $8$ full rotations around the phantom during the same movement cycle. Therefore projections from angles $\theta_i$ and $\theta_{i + 45 k}$ are always taken from the same direction but at different stages of the movement. This data can be viewed as if it was obtained with a faster measurement device or from a slower target.

The movement used in both \texttt{stempo\uns cont360\uns *} and \texttt{stempo\uns seq8x45\uns *} data sets is the same: between each consecutive projection the block moves forward in a straight line by one interleaved step ($\approx 0.18$ mm). Rigorous mathematical formulation of the underlying deformation model $\Phi(\vec{x},t)$ or vector field $\nu$ is not available. Instead it is up to the user to define this information from the data and descriptions using their desired method, format and accuracy, and given the many choices such as spatial and temporal resolution used. However as explained in the next subsection, one approximation of the movement is provided.

\subsection{Complementary data}
\begin{figure}[!bth]
    \centering
    \setlength{\tabcolsep}{1pt}
    \begin{tabular}{lccc}
    $t: \hspace{10mm} 1$ & $120$ & $240$ & $360$ \\
    \includegraphics[width=0.24\columnwidth]{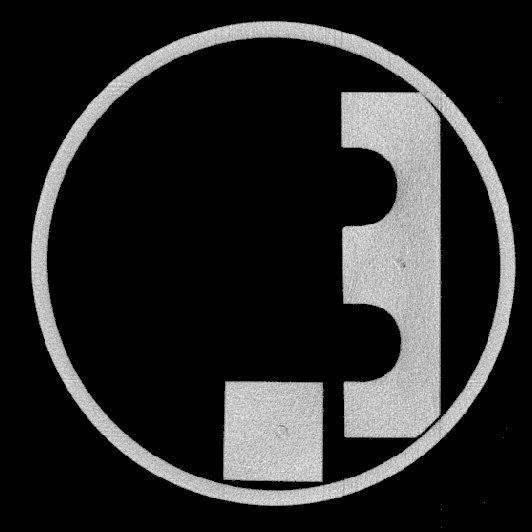} & \includegraphics[width=0.24\columnwidth]{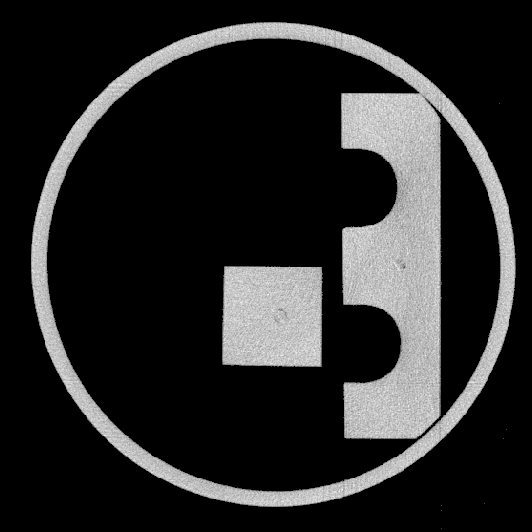} &
    \includegraphics[width=0.24\columnwidth]{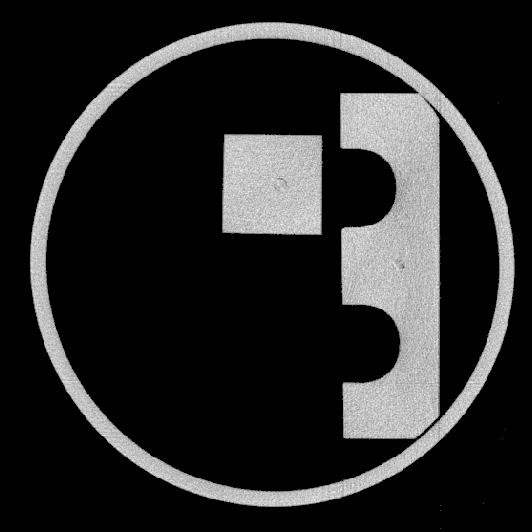} & \includegraphics[width=0.24\columnwidth]{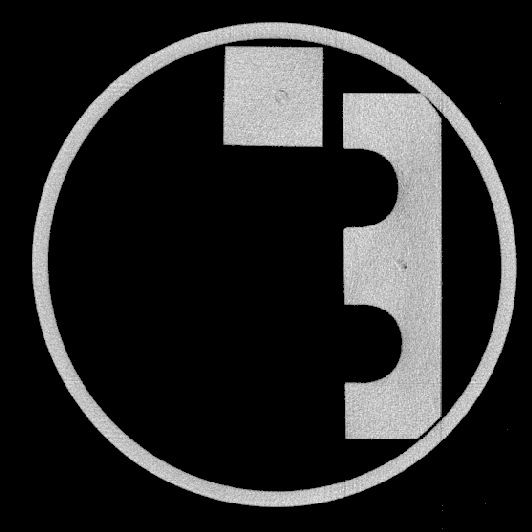} \\
    \end{tabular}
    \caption{Time steps $t = 1,120,240$ and $360$ (out of $360$) of the interpolated ground truth \texttt{obj}.}
    \label{fig:groundTruth}
\end{figure}

The file \texttt{stempo\uns ground\uns truth\uns 2d\uns b4.mat} contains a filtered backprojection (FBP) reconstruction of the static phantom ($560 \times 560$ array \texttt{objStatic}). To reduce the effect of measurement noise, it has been thresholded by removing any pixel values smaller than $1.9 \times 10^{-3}$. This provides a cleaner baseline. This single time step has then been used to manually extrapolate $15$ additional time steps by translating the moving block. These 16 manually obtained translations have then been used to interpolate the new indices corresponding to the translations for all 360 time steps. The resulting images are stored in the $560 \times 560 \times 360$ array \texttt{obj}. Some time steps are shown in Fig.~\ref{fig:groundTruth}. In addition all 360 translations (rounded to an integer to match the row and column indices of the image) are stored in the vectors \texttt{rowShiftInterp} and \texttt{colShiftInterp}.

While \texttt{obj} is inherently affected by errors from the initial reconstruction and the interpolation of the translations, it should be a reasonable approximation of the true motion. Due to the simple nature of the motion, one can easily obtain new approximations if needed.

\textbf{Remark:} downsampling (or upsampling) \texttt{obj} retains the correct scale only if the target resolution is equal to the number of detector elements $D = \tfrac{2240}{\text{binning}}$. For example if a $300 \times 300$ ground truth image is needed to match a $300 \times 300$ reconstruction, \texttt{obj} would need to be downsampled to $260 \times 260$ pixels first and then padded to the final size to avoid stretching. This is because changing the reconstruction resolution in ASTRA does not affect the relative pixel size.

\subsection{Forward operators} \label{ssec:operators}
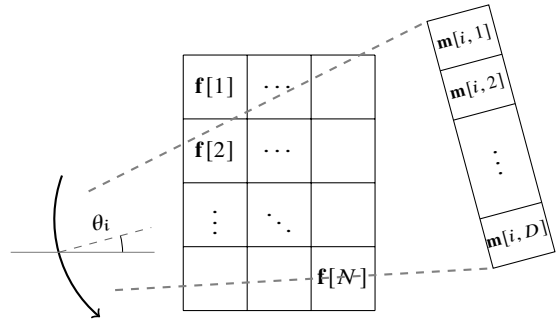
\begin{figure}[hbt]
    \centering
    \begin{tikzpicture}[scale=0.85]
    \begin{scope}[shift={(0.5,-1.3)}]
        \draw[step=1.0,black,thin] (0,0) grid (3,4);
        \node at (0.5,3.5) {$\vec{f}[1]$};
        \node at (0.5,2.5) {$\vec{f}[2]$};
        \node at (0.5,1.5) {$\vdots$};
        \node at (1.5,3.5) {$\dots$};
        \node at (1.5,2.5) {$\dots$};
        \node at (1.5,1.5) {$\ddots$};
        \node at (2.5,0.5) {$\vec{f}[N]$};
    \end{scope}
    
    \begin{scope}[rotate=15]
        \coordinate (C) at (0,0); 
        \draw[gray, dashed, thick] ($(C) + (-.8,.8)$) -- ($(C) + (5,2)$);
        \draw[gray, dashed, thick] ($(C) + (-.8,-.8)$) -- ($(C) + (5,-2)$);

        \draw[->, black, thick] ($(C) + (-1.5,0)$) arc (180:215:2cm);
        \draw[black, thick] ($(C) + (-1.5,0)$) arc (180:145:2cm);
        \draw[thin, black] ($(C) + (-.5,0)$) arc (0:-15:1);
        \draw[gray, thin] ($(C) + (-1.5,0)$) -- ++ (-15:1.5);
        \draw[gray, thin] ($(C) + (-1.5,0)$) -- ++ (165:.75);
        \draw[black!60!white, dashed] ($(C) + (-1.5,0)$) -- ++ (1.5,0) node[midway, above, black, rotate=15] {$\Large{\theta_i}$};
    
        \coordinate (D) at (5.5,0); 
        \draw[black,thin] ($(D) + (-.5,2)$) rectangle ($(D) + (.5,-2)$);
        \draw[black,thin] ($(D) + (-.5,1.2)$) -- ++(1,0);
        \draw[black,thin] ($(D) + (-.5,0.4)$) -- ++(1,0);
        \draw[black,thin] ($(D) + (-.5,-1.2)$) -- ++(1,0);
        \node[rotate=15] at ($(D) + (0,1.6)$) {\scalebox{0.8}{$\vec{m}[i,1]$}};
        \node[rotate=15] at ($(D) + (0,.8)$) {\scalebox{0.8}{$\vec{m}[i,2]$}};
        \node[rotate=15] at ($(D) + (0,-.3)$) {$\vdots$};
        \node[rotate=15] at ($(D) + (0,-1.6)$) {\scalebox{0.8}{$\vec{m}[i,D]$}};
    \end{scope}
    
    \end{tikzpicture}
    \caption{Simplified illustration of how the 2D projections are modelled.} \label{fig:measmodel}
\end{figure}

For an $N = N_x \times N_y$ (in 2D) or $N = N_x \times N_y \times N_z$ (in 3D) target $\vec{f}$ defined by its non-negative attenuation values $\vec{f}[j],\;j = 1,2,...,N$, the mathematical model for the CT measurements can be expressed as
\[
A\vec{f(:)} = \vec{m(:)},
\]
where $A$ is the forward operator and $(:)$ indicates that the vectors $\vec{f}$ and $\vec{m}$ are considered as column vectors. For $P$ projections and $D$ (or $D_r \times D_c$) detector elements, $A$ can be expressed as a sparse $PD \times N$ matrix. The values of $A$ are determined by which pixels of $\vec{f}$ are intersected by a ray from angle $\theta_i$ on the path to a detector element $j$. Figure~\ref{fig:measmodel} illustrates how a single projection is formed in 2D.

The metadata allows for easy construction of the forward operators based on the ASTRA Toolbox. As the dimensions $P, D$ and $N$ get bigger, the operators become computationally very expensive even if sparse matrices are used. Therefore the use of Spot operators is highly recommended (or in the case of 3D geometry, mandatory). These computationally efficient operators allow for some matrix-like operations to be performed, including multiplication by scalars, vectors and matrices, transpose and Kronecker product for building block-diagonal operators. However Spot operators can not be saved into \texttt{.mat}-files and therefore have to be constructed on-the-fly by the user.

A MATLAB function for creating the operator on a CPU is provided by the HelTomo Toolbox as \texttt{create\uns ct\uns operator\uns 2d\uns fan\uns astra.m}. Computers with CUDA compatible graphics cards can also create a faster, GPU-adapted version of the operator with \texttt{create\uns ct\uns operator\uns 2d\uns fan\uns astra\uns cuda.m}. For 3D geometry only the CUDA version is available. The use of HelTomo Toolbox is not mandatory since it is nothing but an interface to the ASTRA Toolbox, designed to work with X-ray data measured at the University of Helsinki. However it automates many of the processes which are necessary for accurate modeling of the forward operator and usually do not need to be changed by the user once correctly set up.

While the reconstruction resolution $N = N_x \times N_y \times N_z$ can be freely chosen by the user, it is recommended to choose $N_x = N_y \approx D_c$. If the resolution is too small, outer edges of the target can be left out and if the resolution is too large, unnecessary computations need to be performed. For 3D geometry choosing the vertical resolution $N_z$ is a trickier problem because the conical X-ray beam leads to reconstruction errors at the top and bottom slices of the volume. 

\section{Measurements} \label{sec:measurements}

\begin{figure}
    \sidecaption[t]
    \begin{tikzpicture}
        \node (img) at (0,0) {\includegraphics[width=0.6\columnwidth]{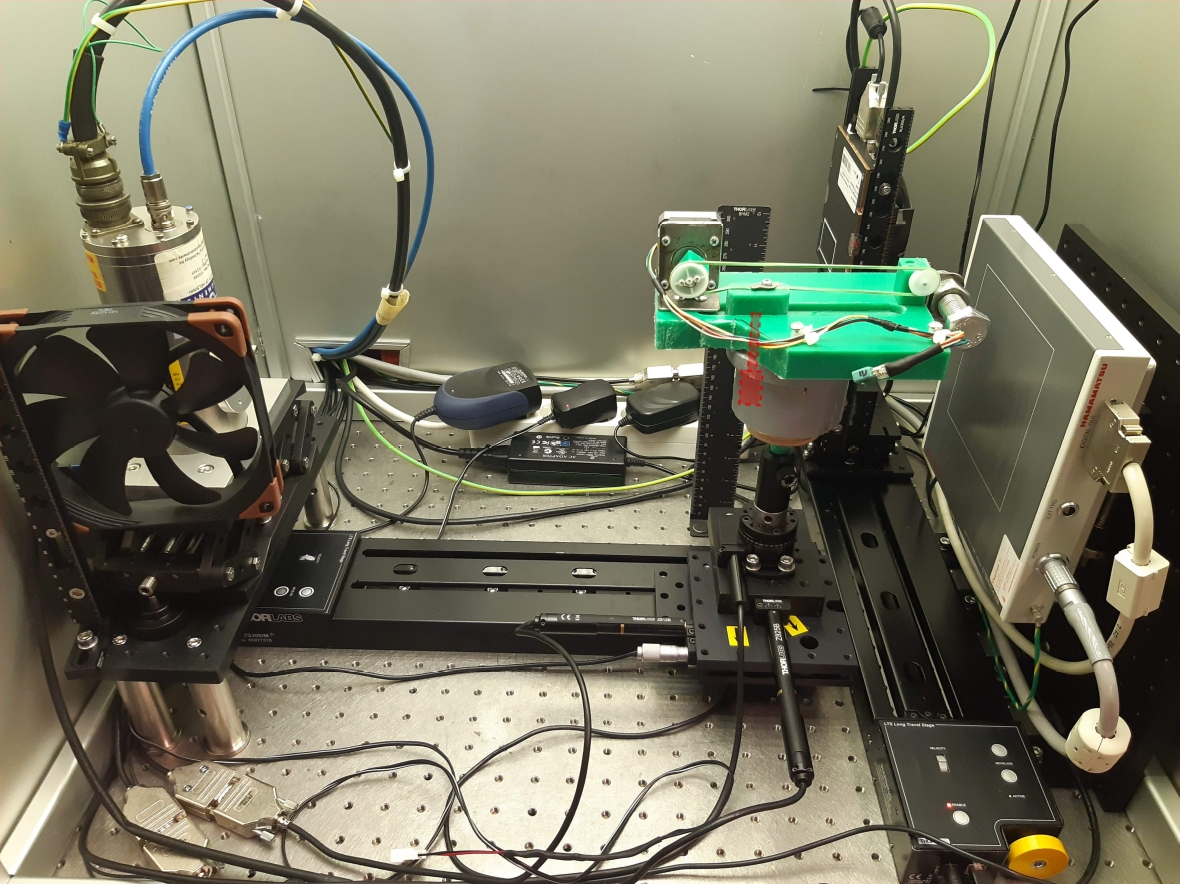}};
        \node[labl] (source) at (-2.7,1) {\bkc{a}}; 
        \node[labl] (det) at (2.6,0.2) {\bkc{b}}; 
        \node[labl] (rot) at (1.1,-0.5) {\bkc{c}}; 
        \node[labl] (phantom) at (1.3,0.6) {\bkc{d}}; 
    \end{tikzpicture}
    \caption{\rev{STEMPO phantom inside the micro-CT device showcasing a: the X-ray source, b: the flat panel detector element, c: the rotating platform and d: the STEMPO phantom with an added illustration of the dynamic object of interest (red).}}
    \label{fig:labels}
\end{figure}
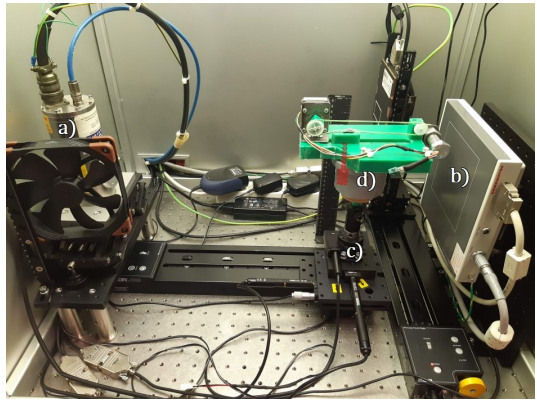

The measurements were done with a scanner designed and constructed in-house in the Industrial Mathematics Computed Tomography Laboratory at the University of Helsinki (original documentation \cite{meaney2015}, later slightly updated). The scanner consists of a molybdenum target X-ray tube (Oxford Instruments XTF5011), a motorized rotation stage (Thorlabs CR1-Z7), and a 12-bit, 2240x2368 pixel, energy-integrating flat panel detector (Hamamatsu Photonics C7942CA-22) with pixel size of $0.05 \times 0.05$ mm. The setup is shown in Fig.~\ref{fig:labels}.

An exposure time of 1000 ms, an X-ray tube acceleration voltage of 40 kV, and a tube current of 1000 mA were used throughout the experiment. Both the rotating platform and the phantom were stationary when the individual projection images were collected. Therefore any motion artifacts are caused by the differences between consecutive projections only.

\begin{figure}[!htb]
    \centering
    \begin{tikzpicture}
    \begin{scope}[scale=0.7]
    
    \coordinate (O) at (2,2);
    \fill[black] (O) circle (2pt) node[below] {Origin};
    
    \coordinate (C) at (-6,2); 
    \draw[gray, dashed, thick] (C)-- ($(C) + (11,2.25)$);
    \draw[gray, dashed, thick] (C) -- ($(C) + (11,-2.25)$);
    
    \filldraw[black, fill=gray!30!white] (C) -- ($(C) + (0.6,0.2) $) -- ($(C) + (0.6,-0.2)$) -- cycle;
    \filldraw[black, fill=gray!30!white] (C) circle (0.45cm);
    \node[right] at ($(C) + (-0.6,-.8)$) {X-ray source};
    
    \filldraw[black, fill=gray!30!white] ($(C) + (10.8,2.3)$) rectangle ($(C) + (11.4,-2.3)$);
    \node[rotate=90] at ($(C) + (11.1,0)$) {Detector};
    
    \draw[|<->|] ($(C) + (0,0.6)$) -- ($(O) + (0,0.6)$) node[midway, below] {SOD};
    \draw[|<->|] ($(C) + (10.8,1.2)$) -- ($(C) + (0,1.2)$) node[midway, above] {SDD};
    \end{scope}
    \end{tikzpicture}
    \caption{Top-down view of the projector geometry (not in scale).} \label{fig:geometry}
\end{figure}
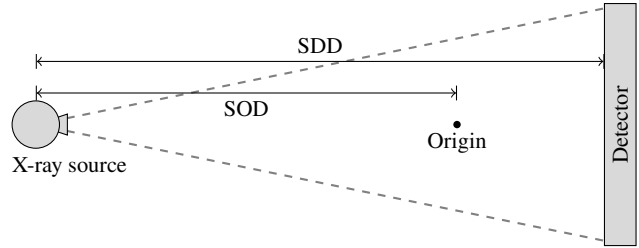

The distance from the X-ray source to the origin (SOD), which acts as the center of rotation, is 410.66~mm, and the distance from the source to the detector (SDD) is 553.74~mm. The conical X-ray beam causes a geometric magnification which needs to be accounted for, but the geometry is stored in the metadata and is automatically handled by the HelTomo and ASTRA Toolboxes. These dimensions and the detector geometry are also illustrated in Fig.~\ref{fig:geometry}.

With this measurement setup, the target is placed on a computer controlled rotational platform which turns clockwise between projections, corresponding to the source-detector pair rotating counter-clockwise around the target.

\section{Example codes} \label{sec:examples}

In addition to the actual data few example codes are provided which should illustrate how the data can be used in practice and in particular how the projections can be subsampled by the user. The examples are not meant to showcase the highest reconstruction quality or computational efficiency but they may give a good baseline for what one can expect from the data.

\begin{itemize}
    \item \texttt{stempo\uns fbp\uns example.m} illustrates the use of the industry-standard FBP algorithm using any of the data and variety of angular subsampling schemes. As expected the method is most reliable when static data and dense angular sampling are used. This example works both with and without CUDA.
    
    \item \texttt{stempo\uns fdk\uns example.m} is similar to the FBP example but uses the full 3D cone-beam geometry and the industry standard FDK (Feldkamp-Davis-Kress) algorithm. It provides 3D volume reconstructions such as the one shown in Fig.~\ref{fig:fdkRecn}. This method is most reliable when static data and dense angular sampling \rev{are} used. This example only works with CUDA.
    \begin{figure}[ht!]
        \sidecaption[t]
        \includegraphics[width=70mm]{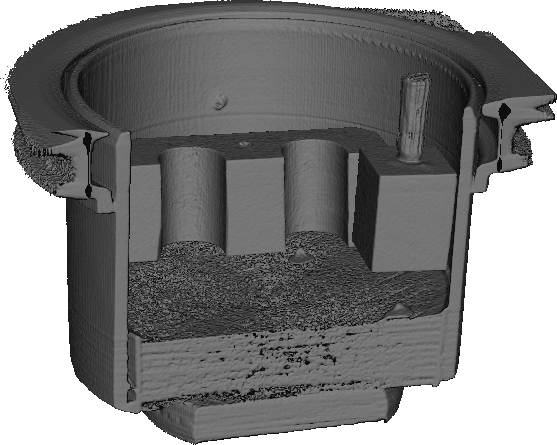}
        \caption{FDK reconstruction of \texttt{stempo\uns static\uns 3d\uns b8} data using 360 projections. Isosurface rendering of a clipped volume for a better view of the interior parts.}
        \label{fig:fdkRecn}
    \end{figure}
    
    \item \texttt{stempo\uns pdfp\uns wavelet\uns 2d\uns example.m} is an iterative reconstruction method which aims at minimizing the functional
    \begin{align*}
        F_{\text{PDFP}}(\vec{f}) := \frac{1}{2} \| A\vec{f} - \vec{m} \|_2^2 + \alpha \| W \vec{f} \|_1,
    \end{align*}
    where $\vec{f}$ is a $N_x \times N_y \times T$ object or animation with non-negative values dropped into a column vector, $A$ is a block diagonal matrix or operator built from discretized Radon transforms for each time step $t = 1, \hdots, T$ and $\vec{m} = [\vec{m}_1, \hdots, \vec{m}_T]^T$ is the stack made of corresponding data or sinograms dropped into a column vector. To enforce connection between time steps the regularization term promotes $\ell^1$-sparse wavelet representation of the whole 3D (2D + time) object using given discrete wavelet transform $W$. The regularization parameter $\alpha$ is automatically tuned as in \cite{purisha2017} and a more detailed explanation of the whole method can be found in \cite{bubba2020}.
    
    This method produces great results with the sequential data if the number of projections is chosen in a balanced fashion. In particular splitting the problem into 16 sets of 23 projections (with some overlap) as in the example code seems to avoid most motion and limited angle artifacts. This can be seen in Fig.~\ref{fig:pdfpHaar}. This example works without CUDA but $A$ can easily be replaced with a GPU-adapted variant.
    
    The computational walltime of the example code was 1182~s for 499 iterations (2.37~s per iter.) on a PC with 16~GB of 2.40~GHz DDR4 memory and Intel i5 CPU at 2.80~GHz.
    
    \begin{figure}[ht!]
        \centering
        \setlength{\tabcolsep}{1pt}
        \begin{tabular}{lccc}
        $t: \hspace{10mm} 1$ & $6$ & $11$ & $16$ \\
        \includegraphics[width=0.24\columnwidth]{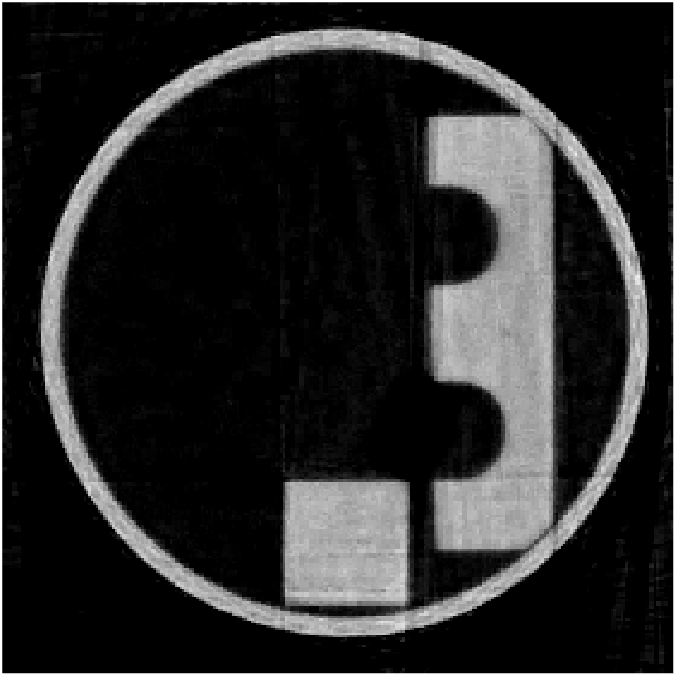} & \includegraphics[width=0.24\columnwidth]{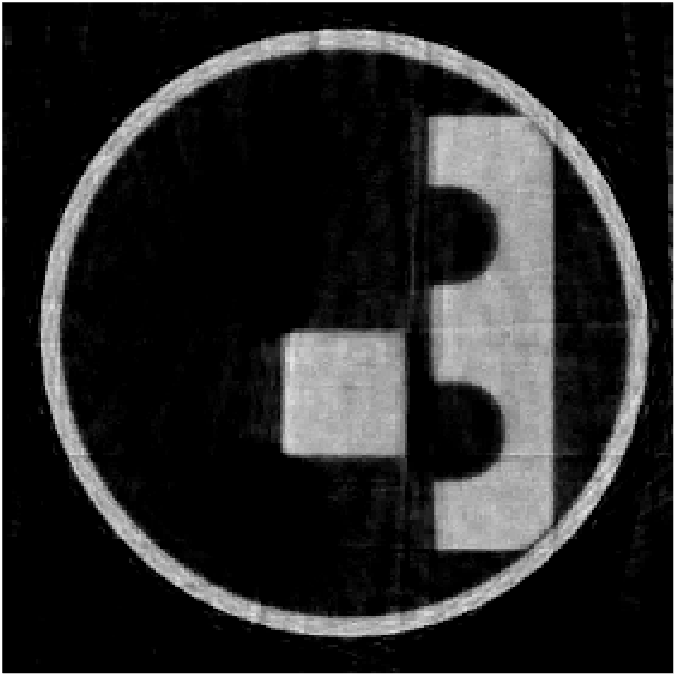} &        \includegraphics[width=0.24\columnwidth]{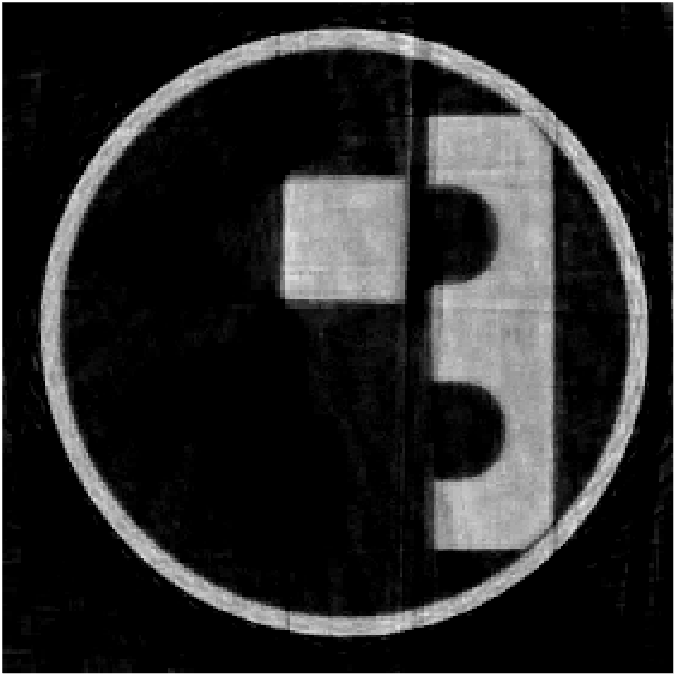} & \includegraphics[width=0.24\columnwidth]{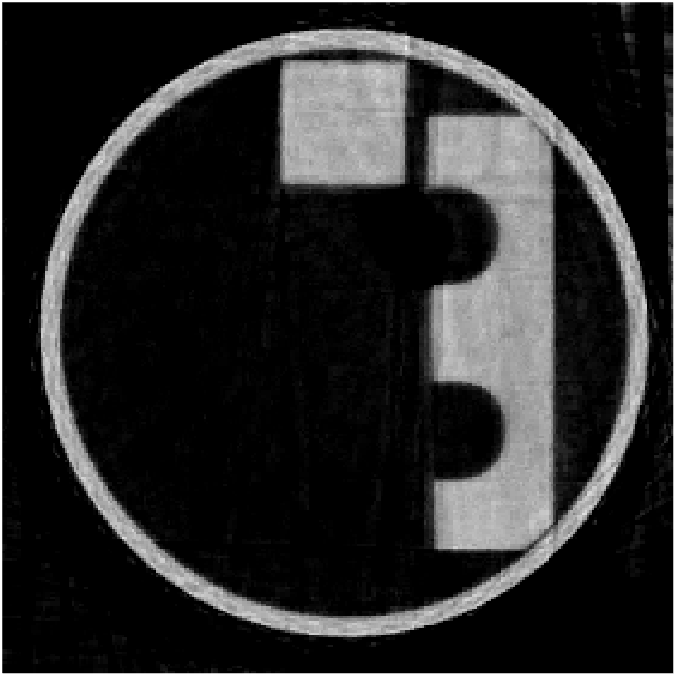} \\
        \end{tabular}
        \caption{Iterative reconstruction of \texttt{stempo\uns seq8x45\uns 2d\uns b8} data using \texttt{stempo\uns pdfp\uns wavelet\uns 2d\uns example.m} with 3D Haar wavelets. Time steps $t = 1,6,11$ and $16$ (out of $16$) are shown.}
        \label{fig:pdfpHaar}
    \end{figure}
    
    \item \texttt{stempo\uns LplusS\uns 2d\uns example.m} is an iterative reconstruction method which aims at separating the slowly changing low rank background $L$ and the more rapidly changing dynamic component $S$ which we expect to be sparse (at least with respect to some representation).
    
    This is obtained by minimizing the regularized functional
    \begin{align*}
        F_{\text{L+S}}(L, S) := \frac{1}{2} \| A(L + S) - \vec{m} \|_2^2 + \mu_L \| L \|_* + \mu_S \| \tilde{W} S \|_1,
    \end{align*}
    where $L$ is the low-rank component formulated as a tall matrix (also denoted $L$) whose columns are the different time steps, $S$ is the sparse component, $A$ is again a block diagonal forward operator and $\tilde{W}$ is a 2D wavelet transform which we compute on every time step of $S = [S_1, \dots, S_T]$. $\| L \|_*$ is the nuclear norm given by the sum of singular values of $L$ and $\mu_L$, $\mu_S$ are the regularization parameters for the respective low-rank and sparse regularization terms. The algorithm is adapted from \cite{otazo2015} by replacing the forward operator in the data mismatch and the Fourier transform with a wavelet transform in the sparsity term. This approach was first applied to 4D CT \cite{gao2011} and later to dynamic PET (positron emission tomography) \cite{yu2015} using framelets.
    
    This method works well with the sequential data if the number of time steps is sufficiently large, as that directly affects the number of columns in $L$ and hence the number of singular values. In the example code 85 time steps are used by considering 24 projections per time step and overlapping the angles in the next time step such that only 4 new projections are used. Illustrations of the $L$ and $S$ components as well as their sum are shown in Fig.~\ref{fig:LplusS}. This example works without CUDA but $A$ can easily be replaced with a GPU-adapted variant.
    
    The computational walltime of the example code was 446~s for 119 iterations (3.75~s per iter.) on a PC with 16~GB of 2.40~GHz DDR4 memory and Intel i5 CPU at 2.80~GHz.
    
    \begin{figure}[ht!]
        \centering
        \setlength{\tabcolsep}{1pt}
        \begin{tabular}{rcccc}
        $t:$ & $1$ & $29$ & $57$ & $85$ \\
         \rotatebox{90}{\hspace{3.4em} $L + S$}\hspace{2pt} & \includegraphics[width=0.23\columnwidth]{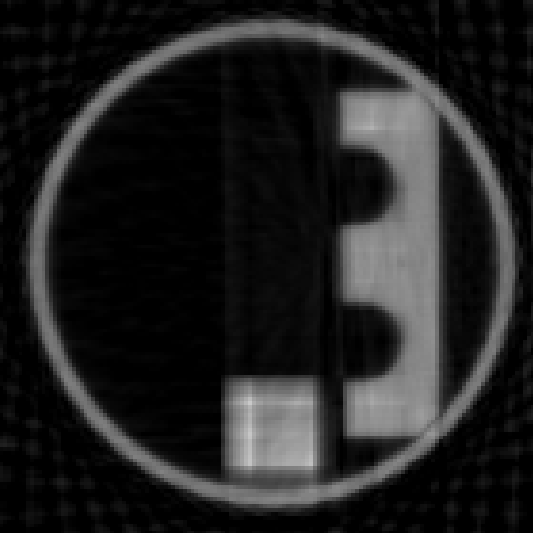} & \includegraphics[width=0.23\columnwidth]{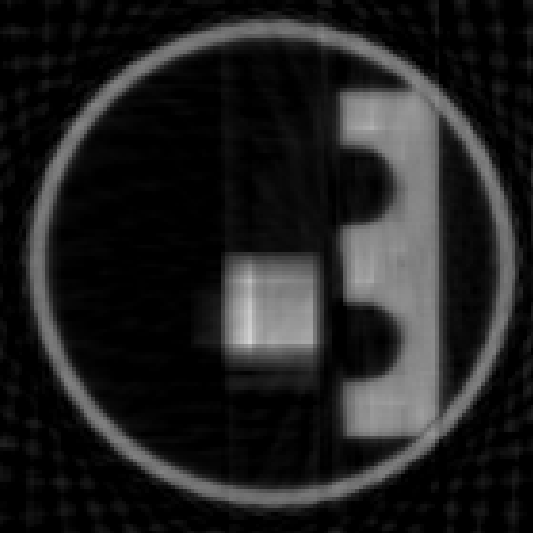} & \includegraphics[width=0.23\columnwidth]{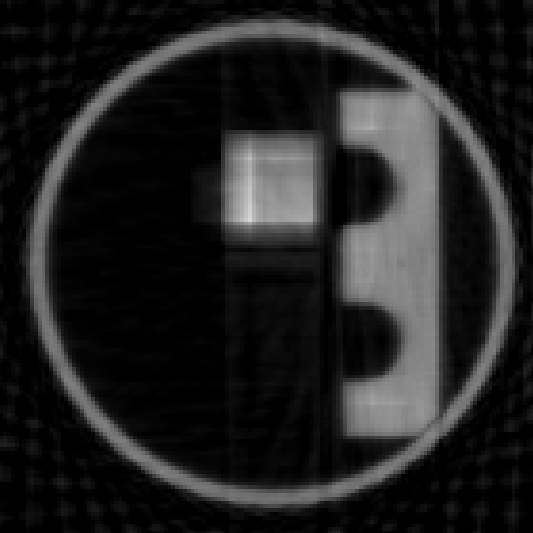} & \includegraphics[width=0.23\columnwidth]{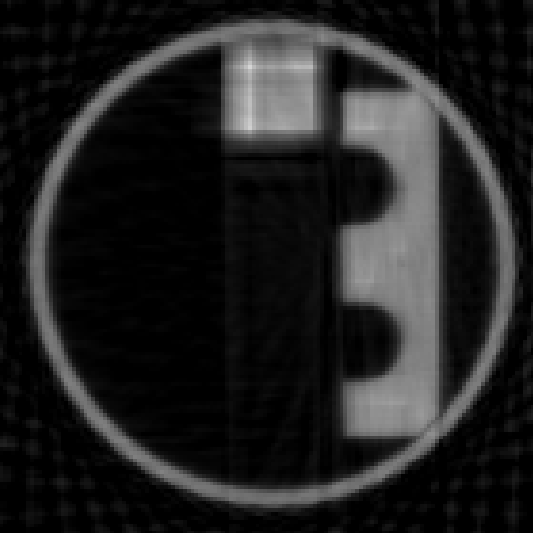} \\
        \rotatebox{90}{\hspace{4.3em} $L$}\hspace{2pt} & \includegraphics[width=0.23\columnwidth]{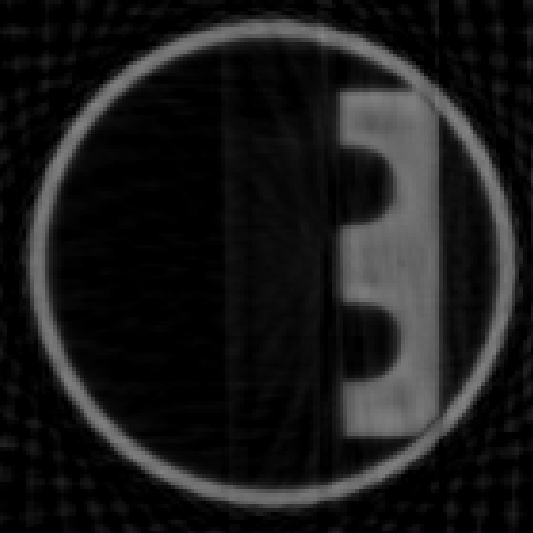} & \includegraphics[width=0.23\columnwidth]{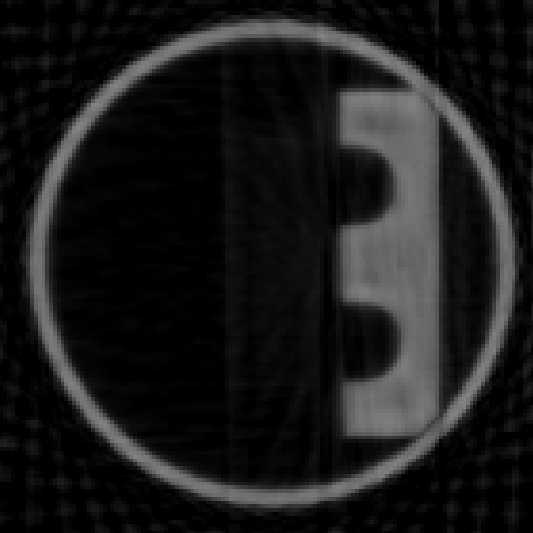} & \includegraphics[width=0.23\columnwidth]{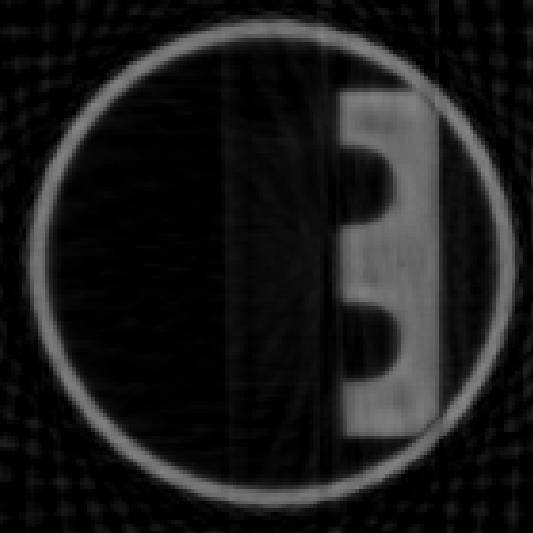} & \includegraphics[width=0.23\columnwidth]{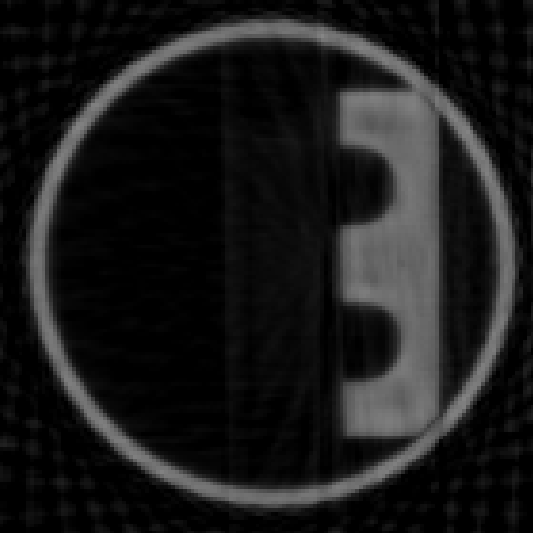} \\ \rotatebox{90}{\hspace{4.3em} $S$}\hspace{2pt} & \includegraphics[width=0.23\columnwidth]{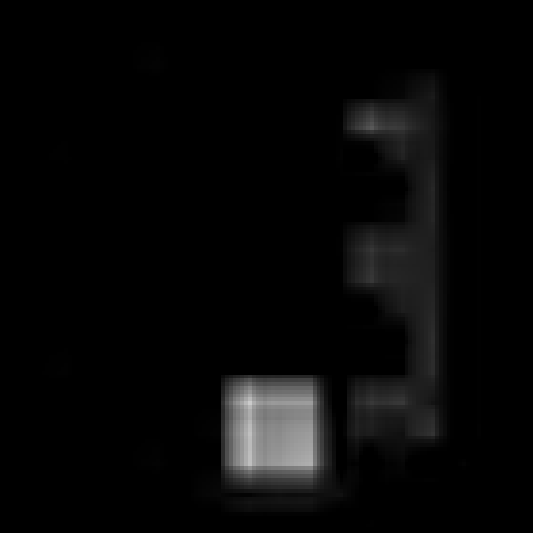} & \includegraphics[width=0.23\columnwidth]{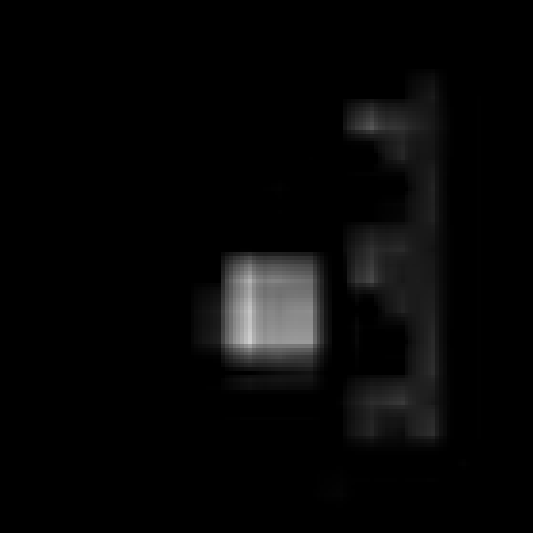} & \includegraphics[width=0.23\columnwidth]{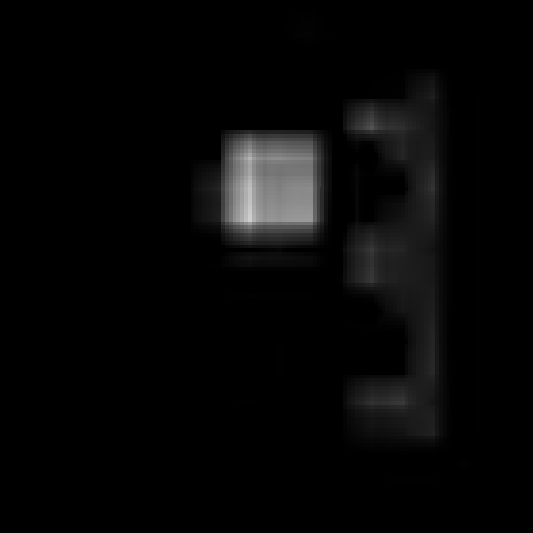} & \includegraphics[width=0.23\columnwidth]{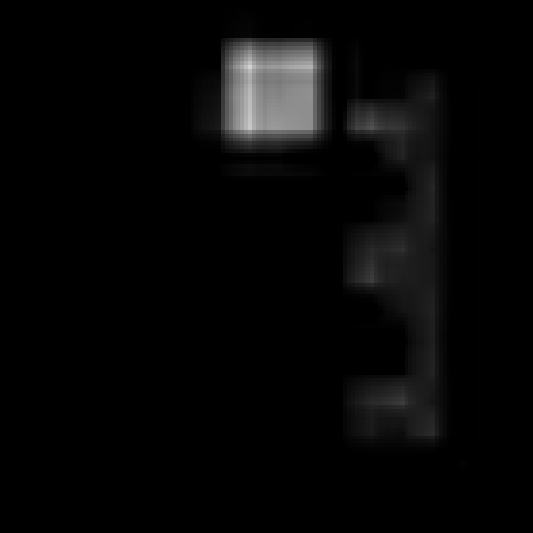} \\
        \end{tabular}
        \caption{Iterative reconstruction of \texttt{stempo\uns seq8x45\uns 2d\uns b8} data using \texttt{stempo\uns LplusS\uns 2d\uns example.m} with 2D Daubechies wavelets. Time steps $t = 1,29,58$ and $85$ (out of $85$) are shown.}
        \label{fig:LplusS}
    \end{figure}
\end{itemize}

\section{Future development} \label{sec:development}

One of the goals of STEMPO is to be flexible and modular enough (both in structure and use) to allow for a wider variety of additional measurements to be collected in the future. Some examples are outlined in this section.

Instead of moving the dynamic block at a constant rate in one direction only, the system easily allows to change the speed and vary the direction (along the fixed channel). For some methods or applications a periodic motion would perhaps be more fitting (e.g. \cite{ford2003}). Alternatively the block could remain mostly static but rapidly change position in the middle of the measurement cycle to simulate sudden unwanted shifting for studying the resulting artifacts.

While the space inside the bounding tube is limited, both the moving block and the static detail can be removed or replaced with different objects with similar attenuation properties. The current choices are homogeneous and there is relatively little structural information to reconstruct from the moving block for example. In the future it may be necessary to make the task more difficult and require greater accuracy from the reconstruction methods by using more complicated target. The sharp corners may also produce too many unwanted line artifacts with some methods.

Finally there is the sampling of the projection directions which offers many interesting choices. For example random and interlaced sampling schemes have been used in the past (cf. \cite{burger2017, bubba2022shearlet, mohan2015}) but neither is easily implemented with the data provided here, since the available projection directions are limited to one per time step. A dataset containing each time step from multiple directions would ease random sampling. This would also offer more training data for any machine learning applications (e.g. \cite{tan2015tensor} or \cite{hauptmann2021} and the references therein) and allow for imitating multi-source systems (cf. \cite{niemi2015}) by repeating the same movement but shifting the projection angles by a fixed amount (such as $90^\circ$). Collecting such data with the current setup would be relatively easy but notably more time consuming especially if very high redundancy is desired.

Some limitations of STEMPO phantom are also worth mentioning. As the phantom (and most current reconstruction methods) are aimed for 2D geometry, allowing vertical movement is a nontrivial task as it would most likely require changes to the lid and motor actuated components. Similarly changing the shape of the channel or some other method of allowing new trajectories is possible but not in the plans in the near future. Even though it is possible to step the motor at just 1/16th the current speed by enabling micro-stepping, very fine movement is not possible since with such slow speeds the elastic pulley (i.e. rubber band) and the friction between the lid and the sled might lead to too inconsistent movement.

Finally some of the design choices (such as the shape, size, cables and attachment method) of this particular phantom may limit its use in other tomographic scanners, if for example higher acceleration voltages were needed in the future. For example the phantom is too wide for our photon counting detector ruling out spectral data in the near future. However the general concept of the phantom should extend relatively easily to other imaging setups and perhaps even to other imaging methods. In fact very recently a similar construction was used in a rigid body motion imaging experiment \cite{djurabekova2022}. Any improvements and new ideas around this topic are always welcome.

\section*{Acknowledgements}
The author is supported by the Emil Aaltonen Foundation junior researcher grant no. 200029 and the Vilho, Yrjö and Kalle Väisälä Foundation of the Finnish Academy of Science and Letters. The author also acknowledges the support of Academy of Finland through the Finnish Centre of Excellence in Inverse Modelling and Imaging 2018–2025, decision number 312339. Finally the author would like to thank E.~Heikkilä, T.~Heikkilä, A.~Meaney and F.S.~Moura for all their technical expertise and help in developing, building and imaging the mechanism.

\printbibliography

\end{document}